\documentclass[showpacs,showkeys,amsmath,amssymb,preprint]{revtex4}
\usepackage{epsfig,dcolumn,bm}

\begin{document}

\title{Coupled-channels study of $\Lambda K$ and $\Sigma K$ states in the chiral
SU(3) quark model}

\author{F. Huang$^{1,2,3}$}
\author{D. Zhang$^{2,3}$}
\author{Z.Y. Zhang$^2$}
\author{Y.W. Yu$^2$}
\affiliation{\small
$^1$CCAST (World Laboratory), P.O. Box 8730, Beijing 100080, PR China \\
$^2$Institute of High Energy Physics, P.O. Box 918-4, Beijing 100049, PR China\footnote{Mailing address.} \\
$^3$Graduate School of the Chinese Academy of Sciences, Beijing, PR China
}

\date{April 24, 2005}

\begin{abstract}
The $S$-wave $\Lambda K$ and $\Sigma K$ states with isospin
$I=1/2$ are dynamically investigated within the framework of the
chiral SU(3) quark model by solving a resonating group method
(RGM) equation. The model parameters are taken from our previous
work, which gave a good description of the energies of the baryon
ground states, the binding energy of the deuteron, and the
experimental data of the nucleon-nucleon ($NN$) and
nucleon-hyperon ($NY$) scattering. Assumed not to give important
contributions in the scattering processes, the $s$-channel
quark-antiquark ($q\bar q$) annihilation interactions are not
included as a first step. The results show a strong attraction
between $\Sigma$ and $K$, which consequently results in a $\Sigma
K$ quasi-bound state with about $17$ MeV binding energy, unlike
the case of $\Lambda K$ which is unbound. When the channel
coupling of $\Lambda K$ and $\Sigma K$ is considered, a sharp
resonance state near $1670$ MeV with spin-parity $J^P=1/2^-$ is
found. The narrow gap of the $\Lambda K$ and $\Sigma K$
thresholds, the strong attraction between $\Sigma$ and $K$, and
the sizeable off-diagonal matrix elements of $\Lambda K$ and
$\Sigma K$ are responsible for the appearance of this resonance.
\end{abstract}



\preprint{hep-ph/0501102}

\maketitle

\section{Introduction}

Recently the BES Collaboration observed a significant threshold
enhancement in the $\Lambda K$ invariant mass spectrum from
$J/\psi\rightarrow K^+\Lambda \bar p$ decays
\cite{sjin04,xyshen04,sjin04c}. The mass of this threshold
structure $N^*_X$ is in the range of 1500 MeV to 1650 MeV, the
width is about $70-110$ MeV, and its spin-parity favors $1/2^-$.
This threshold structure has a very large coupling to the $\Lambda
K$ final state. It is argued to be neither $N^*(1535)$ because of
the non-observation of this state in the SAPHIR experiment $\gamma
p\rightarrow\Lambda K^+$ \cite{glander04}, nor $N^*(1650)$ since
the branching ratio of $N_X^*\rightarrow\Lambda K^+$ is larger
than $20\%$ [for $N^*(1650)\rightarrow\Lambda K^+$ it's $3-11\%$].
But up to now it has not been confirmed finally. Theoretically, to
obtain a proper understanding and a reasonable interpretation of
this enhancement, getting some information about the $\Lambda K$
interaction is necessary.

The other highlight that attracts our attention to the study of
the $\Lambda K$ system is the nucleon resonance $S_{11}(1535)$, of
which the traditional picture is that of an excited three quark
state,  with one of the three quarks orbiting in an $l=1$ state
around the other two \cite{isgur78,isgur79,glozman96}. However
this scenario has difficulties in explaining the large ($30-55\%$)
$N\eta$ decay branching ratio. Differing from the description in
the constituent quark model (CQM), on the hadron level the
$S_{11}(1535)$ is argued to be a quasi-bound $\Lambda K-\Sigma K$
state based on an investigation using the so-called effective
chiral lagrangian (ECL) approach \cite{kaiser95,kaiser97,inoue02}.
In Ref. \cite{kaiser95}, the authors find a strong attraction
between $\Sigma$ and $K$, and thus a bound state will be
necessarily formed below the $\Sigma K$ threshold. This state has
a strong coupling to the $N\eta$ channel, so it is argued to be
the $S_{11}(1535)$ but not $S_{11}(1650)$. Nevertheless, in Ref.
\cite{schutz98}, the authors conclude that the $S_{11}(1535)$ is
not only generated by coupling to higher channels, but appears to
require a genuine three-quark component. So up to now the physical
nature of the $S_{11}(1535)$
--- whether it is an excited three quark state or a quasi-bound
baryon-meson $S$-wave resonance or a mixing of these two
possibilities --- is still a stimulating problem. A dynamical
study on a quark level of the $\Lambda K$ and $\Sigma K$
interactions will undoubtedly make for a better understanding of
the nucleon resonance $S_{11}(1535)$ and $S_{11}(1650)$.

In spite of great successes, the constituent quark model needs to
give a logical explanation, from the underlying theory of the
strong interaction [i.e., Quantum Chromodynamics (QCD)] of the
source of the constituent quark mass. Thus spontaneous vacuum
breaking has to be considered, and as a consequence the coupling
between the quark field and the Goldstone boson is introduced to
restore the chiral symmetry. In this sense, the chiral quark model
can be regarded as a quite reasonable and useful model to describe
the medium-range nonperturbative QCD effect. By generalizing the
SU(2) linear $\sigma$ model, a chiral SU(3) quark model is
developed to study the systems with strangeness \cite{zyzhang97}.
This model has been quite successful in reproducing the energies
of the baryon ground states, the binding energy of the deuteron,
the nucleon-nucleon ($NN$) scattering phase shifts of different
partial waves, and the hyperon-nucleon ($YN$) cross sections by
performing the resonating group method (RGM) calculations
\cite{zyzhang97,lrdai03}. Inspired by these achievements, we try
to extend this model to study the baryon-meson interaction. In
Refs. \cite{fhuang04nk,fhuang04dk}, we studied the $S$-, $P$-,
$D$-, $F$-wave kaon-nucleon ($KN$) phase shifts and fortunately,
we got quite reasonable agreement with the experimental data. At
the same time, the results also show that the effects of the
$s$-channel quark-antiquark ($q\bar q$) annihilation interactions
can be neglected in the scattering processes, since they act only
in the very short range.

In this work, we dynamically study the $\Lambda K$ and $\Sigma K$
states with isospin $I=1/2$ using our chiral SU(3) quark model by
solving the RGM equation.  All the model parameters are taken from
our previous work \cite{zyzhang97,lrdai03}, which gave a
satisfactory description for the energies of the baryon ground
states, the binding energy of the deuteron, the $NN$ scattering
phase shifts, and the $YN$ cross sections. It is assumed that the
$s$-channel $q\bar q$ interactions do not contribute significantly
to a molecular state or a scattering process which is the subject
of our present work, and thus, as a preliminary study they are not
included in the $\Lambda-K$ and $\Sigma-K$ interactions. Our
numerical results show a strong attraction between $\Sigma$ and
$K$, which is qualitatively consistent with the effective chiral
lagrangian calculation \cite{kaiser95}. A $\Sigma K$ quasi-bound
state is thus consequently formed with a binding energy of about
$17$ MeV while the $\Lambda K$ is unbound. Further the channel
coupling of $\Lambda K$ and $\Sigma K$ is considered and the
$\Lambda K$ scattering phase shifts show a sharp resonance with a
mass $M\approx 1669$ MeV and a width $\Gamma\approx 5$ MeV. The
small mass difference of the $\Lambda K$ and $\Sigma K$
thresholds, the strong attraction between $\Sigma$ and $K$, and
the sizeable off-diagonal matrix elements of $\Lambda K$ and
$\Sigma K$ are responsible for the appearance of this resonance.
Although the effects of the $s$-channel interactions as well as
the coupling to the $N\eta$ and $N\pi$ channels should be
considered further, the present results are interesting and
helpful for understanding the new observation of BES
\cite{sjin04,xyshen04,sjin04c} and the structure of the $S_{11}$
nuclear resonances.

The paper is organized as follows. In the next section the
framework of the chiral SU(3) quark model is briefly introduced.
The results for the $\Lambda K$ and $\Sigma K$ states are shown in
Sec. III, where some discussion is presented as well. Finally, the
summary is given in Sec. IV.

\section{Formulation}

As is well known, nonperturbative QCD effects are very important
in light quark systems.  To consider low-momentum medium-range
nonperturbative QCD effects, an SU(2) linear $\sigma$ model
\cite{gellmann60,fernandez93} is proposed to study the $NN$
interaction. In order to extend the study to systems with
strangeness, we generalized the idea of the SU(2) $\sigma$ model
to the flavor SU(3) case, in which a unified coupling between
quarks and all scalar and pseudoscalar chiral fields is
introduced, and the constituent quark mass can be understood in
principle as the consequence of a spontaneous chiral symmetry
breaking of the QCD vacuum \cite{zyzhang97}. With this
generalization, the interacting Hamiltonian between quarks and
chiral fields can be written as
\begin{eqnarray}\label{hami}
H^{ch}_I = g_{ch} F({\bm q}^{2}) \bar{\psi} \left( \sum^{8}_{a=0}
\sigma_a \lambda_a + i \sum^{8}_{a=0} \pi_a \lambda_a \gamma_5
\right) \psi,
\end{eqnarray}
where $g_{ch}$ is the coupling constant between the quark and
chiral-field, and $\lambda_{0}$ a unitary matrix. $\lambda_{1},
...., \lambda_{8}$ are the Gell-Mann matrix of the flavor SU(3)
group, $\sigma_{0},...,\sigma_{8}$ the scalar nonet fields, and
$\pi_{0},..,\pi_{8}$ the pseudoscalar nonet fields. $F({\bm
q}^{2})$ is a form factor inserted to describe the chiral-field
structure \cite{ito90,amk91,abu91,emh91} and, as usual, it is
taken to be
\begin{eqnarray}\label{faca}
F({\bm q}^{2}) = \left(\frac{\Lambda^2}{\Lambda^2+{\bm
q}^2}\right)^{1/2},
\end{eqnarray}
with $\Lambda$ being the cutoff mass of the chiral field. Clearly,
${H}^{ch}_{I}$ is invariant under the infinitesimal chiral SU(3)
transformation.

From ${H}^{ch}_{I}$, the chiral-field-induced effective
quark-quark potentials can be derived, and their expressions are
given in the following:
\begin{eqnarray}
V_{\sigma_a}({\bm r}_{ij})=-C(g_{ch},m_{\sigma_a},\Lambda)
X_1(m_{\sigma_a},\Lambda,r_{ij}) [\lambda_a(i)\lambda_a(j)] +
V_{\sigma_a}^{\bm {l \cdot s}}({\bm r}_{ij}),
\end{eqnarray}
\begin{eqnarray}
V_{\pi_a}({\bm r}_{ij})=C(g_{ch},m_{\pi_a},\Lambda)
\frac{m^2_{\pi_a}}{12m_{q_i}m_{q_j}} X_2(m_{\pi_a},\Lambda,r_{ij})
({\bm \sigma}_i\cdot{\bm \sigma}_j) [\lambda_a(i)\lambda_a(j)]
+V_{\pi_a}^{ten}({\bm r}_{ij}),
\end{eqnarray}
and
\begin{eqnarray}
V_{\sigma_a}^{\bm {l \cdot s}}({\bm r}_{ij})&=&
-C(g_{ch},m_{\sigma_a},\Lambda)\frac{m^2_{\sigma_a}}{4m_{q_i}m_{q_j}}
\left\{G(m_{\sigma_a}r_{ij})-\left(\frac{\Lambda}{m_{\sigma_a}}\right)^3
G(\Lambda r_{ij})\right\} \nonumber\\
&&\times[{\bm L \cdot ({\bm \sigma}_i+{\bm
\sigma}_j)}][\lambda_a(i)\lambda_a(j)],
\end{eqnarray}
\begin{eqnarray}
V_{\pi_a}^{ten}({\bm r}_{ij})&=&
C(g_{ch},m_{\pi_a},\Lambda)\frac{m^2_{\pi_a}}{12m_{q_i}m_{q_j}}
\left\{H(m_{\pi_a}r_{ij})-\left(\frac{\Lambda}{m_{\pi_a}}\right)^3
H(\Lambda r_{ij})\right\} \nonumber\\
&&\times\left[3({\bm \sigma}_i \cdot \hat{r}_{ij})({\bm \sigma}_j
\cdot \hat{r}_{ij})-{\bm \sigma}_i \cdot {\bm
\sigma}_j\right][\lambda_a(i)\lambda_a(j)],
\end{eqnarray}
with $m_{\sigma_a}$ and $m_{\pi_a}$ for the masses of the scalar
and pseudoscalar mesons, respectively, and
\begin{eqnarray}
C(g_{ch},m,\Lambda)=\frac{g^2_{ch}}{4\pi}
\frac{\Lambda^2}{\Lambda^2-m^2} m,
\end{eqnarray}
\begin{eqnarray}
\label{x1mlr} X_1(m,\Lambda,r)=Y(mr)-\frac{\Lambda}{m} Y(\Lambda
r),
\end{eqnarray}
\begin{eqnarray}
X_2(m,\Lambda,r)=Y(mr)-\left(\frac{\Lambda}{m}\right)^3 Y(\Lambda
r),
\end{eqnarray}
\begin{eqnarray}
Y(x)=\frac{1}{x}e^{-x},
\end{eqnarray}
\begin{eqnarray}
G(x)=\frac{1}{x}\left(1+\frac{1}{x}\right)Y(x),
\end{eqnarray}
\begin{eqnarray}
H(x)=\left(1+\frac{3}{x}+\frac{3}{x^2}\right)Y(x).
\end{eqnarray}

In the chiral SU(3) quark model, besides the chiral-field-induced
quark-quark interaction, which describes nonperturbative QCD
effects for the medium range, to study the baryon structure and
hadron-hadron dynamics, one still needs to include an effective
one-gluon-exchange interaction $V^{OGE}_{ij}$ for the short range,
\begin{eqnarray}
V^{OGE}_{ij}=\frac{1}{4}g_{i}g_{j}\left(\lambda^c_i\cdot\lambda^c_j\right)
\left\{\frac{1}{r_{ij}}-\frac{\pi}{2} \delta({\bm r}_{ij})
\left(\frac{1}{m^2_{q_i}}+\frac{1}{m^2_{q_j}}+\frac{4}{3}\frac{1}{m_{q_i}m_{q_j}}
({\bm \sigma}_i \cdot {\bm \sigma}_j)\right)\right\}+V_{OGE}^{\bm
l \cdot \bm s},
\end{eqnarray}
with
\begin{eqnarray}
V_{OGE}^{\bm l \cdot \bm
s}=-\frac{1}{16}g_ig_j\left(\lambda^c_i\cdot\lambda^c_j\right)
\frac{3}{m_{q_i}m_{q_j}}\frac{1}{r^3_{ij}}{\bm L \cdot ({\bm
\sigma}_i+{\bm \sigma}_j)},
\end{eqnarray}
and a confinement potential $V^{conf}_{ij}$ for the long range,
\begin{eqnarray}
V_{ij}^{conf}=-a_{ij}^{c}(\lambda_{i}^{c}\cdot\lambda_{j}^{c})r_{ij}^2
-a_{ij}^{c0}(\lambda_{i}^{c}\cdot\lambda_{j}^{c}).
\end{eqnarray}

For the systems with an antiquark ${\bar s}$, the total
Hamiltonian can be written as
\cite{fhuang04nk,fhuang04dk,fhuang04theta}
\begin{eqnarray}
\label{hami5q}
H=\sum_{i=1}^{5}T_{i}-T_{G}+\sum_{i<j=1}^{4}V_{ij}+\sum_{i=1}^{4}V_{i\bar
5},
\end{eqnarray}
where $T_G$ is the kinetic energy operator for the center-of-mass
motion, and $V_{ij}$ and $V_{i\bar 5}$ represent the quark-quark
($qq$) and quark-antiquark ($q{\bar q}$) interactions,
respectively,
\begin{eqnarray}
V_{ij}= V^{OGE}_{ij} + V^{conf}_{ij} + V^{ch}_{ij},
\end{eqnarray}
\begin{eqnarray}
V^{ch}_{ij}=\sum^{8}_{a=0}V_{\sigma_a}(\bm
r_{ij})+\sum^{8}_{a=0}V_{\pi_a} (\bm r_{ij}).
\end{eqnarray}
$V_{i \bar 5}$ in Eq. (\ref{hami5q}) includes two parts: direct
interaction and annihilation parts,
\begin{eqnarray}
V_{i\bar 5}=V^{dir}_{i\bar 5}+V^{ann}_{i\bar 5},
\end{eqnarray}
with
\begin{eqnarray}
V_{i\bar 5}^{dir}=V_{i\bar 5}^{conf}+V_{i\bar 5}^{OGE}+V_{i\bar
5}^{ch},
\end{eqnarray}
where
\begin{eqnarray}
V_{i\bar
5}^{conf}=-a_{i5}^{c}\left(-\lambda_{i}^{c}\cdot{\lambda_{5}^{c}}^*\right)r_{i\bar
5}^2
-a_{i5}^{c0}\left(-\lambda_{i}^{c}\cdot{\lambda_{5}^{c}}^*\right),
\end{eqnarray}
\begin{eqnarray}
V^{OGE}_{i\bar
5}&=&\frac{1}{4}g_{i}g_{5}\left(-\lambda^c_i\cdot{\lambda^c_5}^*\right)
\left\{\frac{1}{r_{i\bar 5}}-\frac{\pi}{2} \delta({\bm r}_{i\bar
5})
\left(\frac{1}{m^2_{q_i}}+\frac{1}{m^2_{q_5}}+\frac{4}{3}\frac{1}{m_{q_i}m_{q_5}}
({\bm \sigma}_i \cdot {\bm \sigma}_5)\right)\right\}  \nonumber \\
&&-\frac{1}{16}g_ig_5\left(-\lambda^c_i\cdot{\lambda^c_5}^*\right)
\frac{3}{m_{q_i}m_{q_5}}\frac{1}{r^3_{i\bar 5}}{\bm L \cdot ({\bm
\sigma}_i+{\bm \sigma}_5)},
\end{eqnarray}
and
\begin{eqnarray}
V_{i\bar{5}}^{ch}=\sum_{j}(-1)^{G_j}V_{i5}^{ch,j}.
\end{eqnarray}
Here $(-1)^{G_j}$ represents the G parity of the $j$th meson. The
$q\bar q$ annihilation interactions, $V_{i\bar 5}^{ann}$, are not
included in the $\Lambda-K$ and $\Sigma-K$ interactions in this
preliminary work since they are assumed not to contribute
significantly to a molecular state or a scattering process which
is the subject of our present study.

All the model parameters are taken from our previous work
\cite{zyzhang97,lrdai03}, which gave a satisfactory description
for the energies of the baryon ground states, the binding energy
of the deuteron, the $NN$ scattering phase shifts, and the $YN$
cross sections. Here we briefly give the procedure for the
parameter determination. We have three initial input parameters:
the harmonic-oscillator width parameter $b_u$, the up (down) quark
mass $m_{u(d)}$, and the strange quark mass $m_s$. These three
parameters are taken to be the usual values: $b_u=0.5$ fm,
$m_{u(d)}=313$ MeV, and $m_s=470$ MeV. By some special
constraints, other model parameters are fixed in the following
way. The chiral coupling constant $g_{ch}$ is fixed by
\begin{eqnarray}
\frac{g^{2}_{ch}}{4\pi} = \left( \frac{3}{5} \right)^{2}
\frac{g^{2}_{NN\pi}}{4\pi} \frac{m^{2}_{u}}{M^{2}_{N}},
\end{eqnarray}
with empirical value $g^{2}_{NN\pi}/4\pi=13.67$. The masses of the
mesons are taken to be the experimental values, except for the
$\sigma$ meson. The $m_\sigma$ is treated as an adjustable
parameter and obtained to be 595 MeV by fitting the binding energy
of the deuteron. The cutoff radius $\Lambda^{-1}$ is taken to be
the value close to the chiral symmetry breaking scale
\cite{ito90,amk91,abu91,emh91}. After the parameters of chiral
fields are fixed, the one-gluon-exchange coupling constants
$g_{u}$ and $g_{s}$ are determined by the mass splits between $N$,
$\Delta$ and $\Lambda$, $\Sigma$ respectively. The confinement
strengths $a^{c}_{uu}$, $a^{c}_{us}$, and $a^{c}_{ss}$ are fixed
by the stability conditions of $N$, $\Lambda$, and $\Xi$, and the
zero-point energies $a^{c0}_{uu}$, $a^{c0}_{us}$, and
$a^{c0}_{ss}$ by fitting the masses of $N$, $\Sigma$ and
$\overline{\Xi+\Omega}$, respectively. All the parameters are
tabulated in Table \ref{para}.

{\small
\begin{table}[htb]
\caption{\label{para} Model parameters. The meson masses and the
cutoff masses: $m_{\sigma'}=980$ MeV, $m_{\kappa}=980$ MeV,
$m_{\epsilon}=980$ MeV, $m_{\pi}=138$ MeV, $m_K=495$ MeV,
$m_{\eta}=549$ MeV, $m_{\eta'}=957$ MeV, and $\Lambda=1100$ MeV.}
\begin{center}
\begin{tabular*}{160mm}{@{\extracolsep\fill}ccccc}
\hline\hline
 & $m_u$ (MeV) && 313 &  \\
 & $m_s$ (MeV) && 470 &  \\
 & $b_u$ (fm)  && 0.5 &  \\
 & $g_u$     && 0.886  & \\
 & $g_s$     && 0.917  & \\
 & $m_\sigma$ (MeV) &&  595  & \\
 & $a^c_{uu}$ (MeV/fm$^2$) && 48.1 & \\
 & $a^c_{us}$ (MeV/fm$^2$) && 60.7 & \\
 & $a^c_{ss}$ (MeV/fm$^2$) && 101.2 & \\
 & $a^{c0}_{uu}$ (MeV)  && $-$43.6 & \\
 & $a^{c0}_{us}$ (MeV)  && $-$38.2 & \\
 & $a^{c0}_{ss}$ (MeV)  && $-$36.1 & \\
\hline\hline
\end{tabular*}
\end{center}
\end{table}}

With all parameters determined in the chiral SU(3) quark model,
the $\Lambda K$ and $\Sigma K$ systems can be dynamically studied
in the frame work of the RGM. The wave function of the five quark
system is of the form
\begin{eqnarray}
\Psi=\sum_\beta {\cal A}[{\hat \phi}_A(\bm \xi_1,\bm \xi_2) {\hat
\phi}_B(\bm \xi_3) \chi_{\beta}({\bm R}_{AB})],
\end{eqnarray}
where ${\bm \xi}_1$ and ${\bm \xi}_2$ are the internal coordinates
for the cluster $A$ ($\Lambda$ or $\Sigma$), and ${\bm \xi}_3$ the
internal coordinate for the cluster $B$ ($K$). ${\bm R}_{AB}\equiv
{\bm R}_A-{\bm R}_B$ is the relative coordinate between the two
clusters, $A$ and $B$, and $\beta\equiv (A,B,I,S,L,J)$ specifies
the hadron species ($A,B$) and quantum numbers of the baryon-meson
channel. The ${\hat \phi}_A$ $({\hat \phi}_B)$ is the
antisymmetrized internal cluster wave function of $A$ $(B)$, and
$\chi_\beta ({\bm R}_{AB})$ the relative wave function of the two
clusters. The symbol $\cal A$ is the antisymmetrizing operator
defined as
\begin{equation}
{\cal A}\equiv{1-\sum_{i \in A}P_{i4}}\equiv{1-3P_{34}}.
\end{equation}
Substituting $\Psi$ into the projection equation
\begin{equation}
\langle \delta\Psi|(H-E)|\Psi \rangle=0,
\end{equation}
we obtain the coupled integro-differential equation for the
relative function $\chi_\beta$ as
\begin{eqnarray}\label{crgm}
\sum_{\beta'}\int \left[{\cal H}_{\beta\beta'}(\bm R, \bm
R')-E{\cal N}_{\beta\beta'}(\bm R, \bm R')\right]\chi_{\beta'}(\bm
R') d\bm R' =0,
\end{eqnarray}
where the Hamiltonian kernel $\cal H$ and normalization kernel
$\cal N$ can, respectively, be calculated by
\begin{eqnarray}
\left\{
       \begin{array}{c}
          {\cal H}_{\beta\beta'}(\bm R, \bm R')\\
          {\cal N}_{\beta\beta'}(\bm R, \bm R')
       \end{array}
\right\} =\left<[{\hat \phi}_A(\bm \xi_1,\bm \xi_2 ) {\hat
\phi}_B(\bm \xi_3)]_\beta\delta(\bm R-{\bm R}_{AB})\right.\left|
\left\{
\begin{array}{c}
          H \\
          1
       \end{array}
\right\}
\right| \nonumber \\
\left.{\cal A}\left[[{\hat \phi}_A(\bm \xi_1,\bm \xi_2) {\hat
\phi}_B(\bm \xi_3)]_{\beta'}\delta(\bm R'-{\bm
R}_{AB})\right]\right>.
\end{eqnarray}

Eq. $(\ref{crgm})$ is the so-called coupled-channel RGM equation.
Expanding unknown $\chi_\beta ({\bm R}_{AB})$ by employing
well-defined basis wave functions, such as Gaussian functions, one
can solve the coupled-channel RGM equation for a bound-state
problem or a scattering one to obtain the binding energy or
scattering phase shifts for the two-cluster systems. The details
of solving the RGM equation can be found in Refs.
\cite{kwi77,mka77,mok81,ust88,fhuang04nk}.

\section{Results and discussions}

\begin{figure}[htb]
\epsfig{file=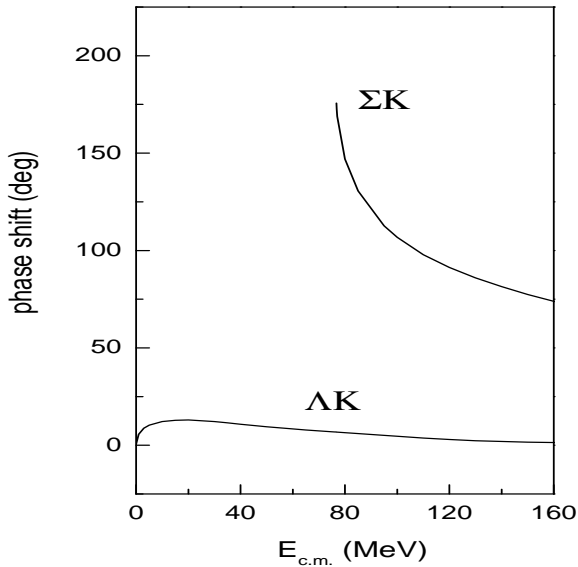,width=8.0cm,height=8.0cm} \vglue -0.5cm
\caption{\small \label{phase} The $S$-wave $\Lambda K$ and $\Sigma
K$ phase shifts in the one-channel calculation.}
\end{figure}

We perform a RGM dynamical study of $\Lambda K$ and $\Sigma K$
states with isospin $I=1/2$ in our chiral SU(3) quark model. Fig.
\ref{phase} shows one-channel-calculation results for the $S$-wave
$\Lambda K$ and $\Sigma K$ ($I=1/2$) elastic scattering phase
shifts. Here the one-channel-calculation means without considering
the channel coupling of $\Lambda K$ and $\Sigma K$. The phase
shifts show up a strong attractive interaction between $\Sigma$
and $K$, which is qualitatively consistent with the effective
chiral lagrangian calculation based on the hadron level
\cite{kaiser95,kaiser97}, and a very weak interaction between
$\Lambda$ and $K$. Our further analysis demonstrates that this
strong attraction between $\Sigma$ and $K$ dominantly comes from
the color magnetic force of OGE and the $\sigma$ meson exchange.
Such a strong attraction can consequently result in a $\Sigma K$
quasi-bound state. A concrete solution of the RGM equation for a
bound state problem shows that $\Sigma K$ is really bound and its
binding energy is about $17$ MeV.  A similar study to the $\Lambda
K$ system is also made, as we expected it is unbound, because
there is no enough attraction between $\Lambda$ and $K$.

\begin{figure}[htb]
\epsfig{file=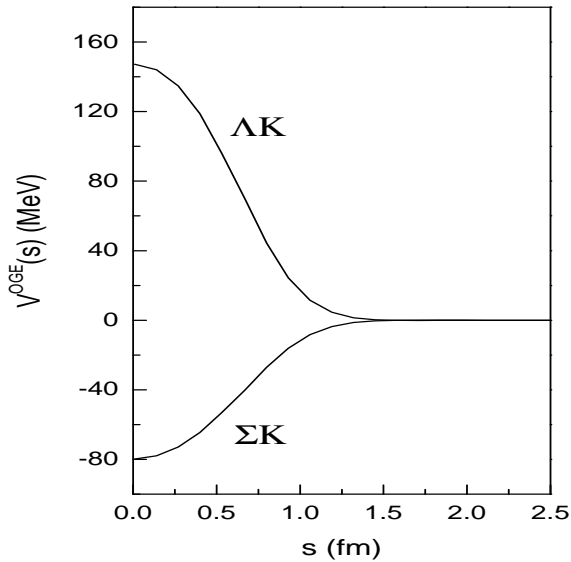,width=8.0cm,height=8.0cm} \vglue -0.7cm
\caption{\small \label{GCM_OGE} The GCM matrix elements of OGE.}
\end{figure}

To understand the results better, an extensive analysis is carried
out to see the contributions from various parts of the
interactions in the $\Lambda K$ and $\Sigma K$ systems. Fig.
\ref{GCM_OGE} shows the diagonal matrix elements of the
one-gluon-exchange potential in the generator coordinate method
(GCM) calculation \cite{kwi77}, which can describe the
interactions between two clusters $\Lambda$ ($\Sigma$) and $K$
qualitatively. In Fig. \ref{GCM_OGE}, $s$ denotes the generator
coordinate and $V^{OGE}(s)$ is the OGE effective potential between
the two clusters. One sees that the OGE effective potential is
attractive for $\Sigma K$ while repulsive for $\Lambda K$. This
property is quite interesting since the attraction of OGE in the
short distance may as a consequence make $\Sigma$ and $K$ to form
a quasi-bound $\Sigma K$ state, and simultaneously the short-range
strong repulsion indicates that it is difficult to form a $\Lambda
K$ bound state. One may argue that the OGE interaction of $NN$
$^3S_1$ partial wave is also repulsive, but a weakly bound
deuteron is also formed. This is understandable if one notices
that the tensor force of the one-pion exchange plays an important
role in reproducing the binding energy of the deuteron
\cite{lrdai03}. But in the $\Lambda K$ and $\Sigma K$ systems the
tensor force totally vanishes since the kaon meson is spin zero
and the total spin of the two clusters is $1/2$. The tensor force
can exist only when $|S-2|=S$, where $S$ is the total spin of two
clusters. This holds for investigations on both quark and hadron
levels. Note on the hadron level there is no one-pion exchange
between $\Lambda$ and $K$ since $\Lambda$ has isospin zero and
$\pi$ has isospin one, while in the quark model study although
comparatively weak the one-pion exchange does exist due to the
quark exchange (see Eqs. 25-26) required by the Pauli principle.

\begin{figure}[htb]
\epsfig{file=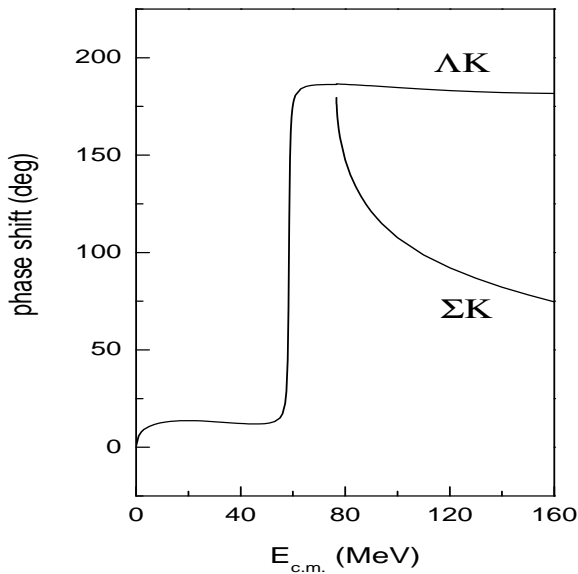,width=8.0cm,height=8.0cm} \vglue
-0.5cm \caption{\small \label{phase_c} The $S$-wave $\Lambda K$
and $\Sigma K$ phase shifts in the coupled-channel calculation.}
\end{figure}

As a second step, we do a two-channel-coupling calculation of
$\Lambda K$ and $\Sigma K$ systems. The phase shifts are shown in
Fig. \ref{phase_c}. One sees that there is a sharp resonance with
a mass $M\approx 1669$ MeV and a width $\Gamma\approx 5$ MeV. It
is not difficult to understand the appearance of this resonance
state if one notices the following three important features. (1)
The mass difference of the $\Lambda K$ and $\Sigma K$ thresholds
is small (about $78$ MeV). Generally speaking, the closer the
thresholds of these two channels are, the larger the
channel-coupling effects could be. (2) There is a strong
attractive interaction between $\Sigma$ and $K$. Such a strong
attraction can result in a $\Sigma K$ quasi-bound state with about
$17$ MeV binding energy. Thus when it couples to the $\Lambda K$
channel, a resonance would be possible to appear between the
thresholds of $\Lambda K$ and $\Sigma K$. (3) The off-diagonal
matrix elements of $\Lambda K$ and $\Sigma K$ are comparatively
big. Such sizeable off-diagonal matrix elements can give a great
impact upon the $\Lambda K$ phase shifts in the coupled-channel
calculation.

\begin{figure}[htb]
\epsfig{file=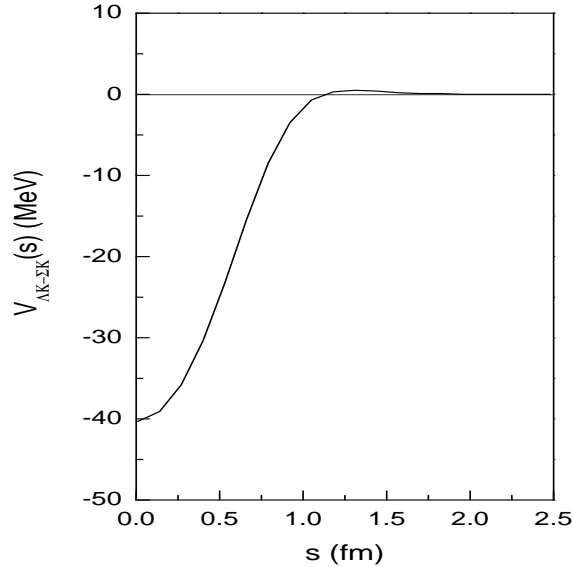,width=8.0cm,height=8.0cm} \vglue
-0.7cm \caption{\small \label{transition} The GCM matrix elements
of the Hamiltonian in the coupled-channel calculation.}
\end{figure}

In the coupled-channel study, the transition potential from
$\Sigma K$ to $\Lambda K$ in our framework is a nonlocal one. In
Fig. \ref{transition} we give the diagonal matrix elements in the
coupled channel of the Hamiltonian in the generator coordinate
method (GCM) calculation, which can describe the transition
potential qualitatively. In this figure, $s$ denotes the generator
coordinate and $V_{\Lambda K-\Sigma K}(s)$ is the effective
transition potential of the coupled channel. One can see that the
matrix elements of the transition interaction from $\Sigma K$ to
$\Lambda K$ are really considerably large. Further analysis
reveals that this interaction dominantly comes from the color
magnetic force of OGE.

\begin{figure}[htb]
\vglue 2.5cm
\epsfig{file=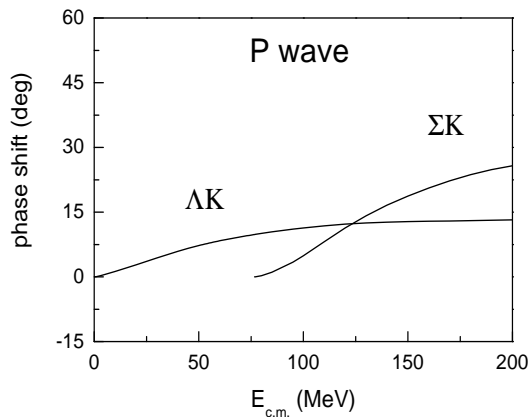,width=8.0cm,height=6.0cm} \vglue
-3.0cm \caption{\small \label{phase_p} The $P$-wave $\Lambda K$
and $\Sigma K$ phase shifts in the coupled-channel calculation.}
\end{figure}

The $P$-wave $\Lambda K$ and $\Sigma K$ phase shifts are also
investigated. Fig. \ref{phase_p} shows the results computed by
solving a two-channel RGM equation. The phase shifts of the
coupled-channel calculation are almost the same as those without
channel coupling. This is reasonable because the $\Lambda K-\Sigma
K$ off-diagonal matrix elements come from the color magnetic force
of OGE while in the $P$ wave such an interaction nearly vanishes.
Thus the channel coupling effect is small enough to be neglected.

One thing should be mentioned that our results are independent of
the confinement potential since between two color-singlet clusters
the confinement potential scarcely contributes any interactions
\cite{fhuang04nk,fhuang04dk}. Thus our numerical results will
almost remain unchanged even the color quadratic confinement is
replaced by the color linear one.

We now discuss the resonance obtained from the coupled-channel
calculation. The present work is just making a dynamical study on
the quark level of the $\Lambda K$ and $\Sigma K$ interactions.
Surely there are other channels that also couple to these two
channels, such as the $N\eta$ and $N\pi$ channels and even a
genuine $3q$ component. These channels are significant and
essential for giving a proper explanation of the states with
$J^P=1/2^-$ and mass between $1500-1700$ MeV ($N_X^*$,
$S_{11}(1535)$, $S_{11}(1650)$). The properties of the $N^*$, such
as that of the $N^*(1535)$ coupling strongly to $N\eta$ and the
$N^*(1650)$ having a large branching ratio to $N\pi$, should be
expected to hold automatically if all the channels were
considered. In this work we only consider the $\Lambda K$ and
$\Sigma K$ channels. Certainly the $N^*(1535)$ can not be
explained since the $N\eta$ channel is omitted. From the mass
point of view and considering that the branching ratio of
$N^*(1650)$ to $\Lambda K$ is $3-11\%$, the resonance we obtained
prefers to be an $N^*(1650)$, though the calculated width is too
small. It is expected that the coupling to $N\pi$ would give rise
to a much larger width than the present calculated one since
empirically the $N\pi$ channel accounts for $55-90\%$ of the width
of $N^*(1650)$. To some extent the possibility of this resonance
to be the $N^*_X$ observed by the BES Collaboration can not be
ruled out since preliminarily the $N^*_X$ has a large branching
ratio ($>20\%$) to the $\Lambda K$ final state, and it's mass is
in the range of $1500$ MeV to $1650$ MeV though up to now they
have not yet been confirmed. Anyhow, the final conclusion
regarding what is this resonance ($N_X^*$, $S_{11}(1535)$,
$S_{11}(1650)$, or any other state) and its exact theoretical mass
and width must wait for further work where more channel couplings
will be included and the decay properties will be studied.

In Refs. \cite{jido03,garcia04}, many low lying resonances are
studied as quasibound meson-baryon states dynamically generated
from the interaction of the octet of pseudoscalar mesons with the
octet of the $1/2^+$ baryons based on investigations using the
effective chiral lagrangian approach. Two octets and one singlet
of dynamically generated resonances are predicted in these works.
The interesting thing for us is that, if the $N^*(1535)$ can
really be explained as a meson-baryon state as claimed by the
authors, then there should be another state with the same quantum
numbers and other properties of the $N^*(1535)$.

\section{Summary}

In summary, we perform a dynamical study of $\Lambda K$ and
$\Sigma K$ states in the framework of the chiral SU(3) quark model
by solving the RGM equation. The model parameters are taken to be
the values determined by the energies of the baryon ground states,
the binding energy of the deuteron, the $NN$ scattering phase
shifts, and the $YN$ cross sections. Because this is a preliminary
study, the $s$-channel $q\bar q$ annihilation interactions are not
included. The results show a strong attraction between the sigma
and kaon, and a $\Sigma K$ quasi-bound state is thus formed as a
consequence with a binding energy of about $17$ MeV, while the
$\Lambda K$ is unbound. When the channel coupling of $\Lambda K$
and $\Sigma K$ is considered, the scattering phase shifts show a
sharp resonance with a mass $M\approx 1669$ MeV and a width
$\Gamma\approx 5$ MeV. The small mass difference of the $\Lambda
K$ and $\Sigma K$ thresholds, the strong attraction between
$\Sigma$ and $K$, and the sizeable off-diagonal matrix elements of
$\Lambda K$ and $\Sigma K$ are responsible for the appearance of
this resonance. The results are interesting and useful for
understanding the new observation of BES
\cite{sjin04,xyshen04,sjin04c} and the structure of the $S_{11}$
nuclear resonances. The final conclusion regarding what is the
resonance we obtained and its exact theoretical mass and width
will wait for further work where the effects of the $s$-channel
$q\bar q$ annihilation as well as the coupling to the $N\eta$ and
$N\pi$ channels and even to a genuine $3q$ component will be
considered and the decay properties will be studied.

\begin{acknowledgements}
One of the authors (F. Huang) is indebted to Prof. B.S. Zou for a
careful reading of the manuscript. This work was supported in part
by the National Natural Science Foundation of China, Grant No.
10475087.
\end{acknowledgements}

\end{document}